\documentclass{moriond}

\usepackage{amsmath}
\usepackage{tikz}
\usetikzlibrary{calc,decorations.pathreplacing,decorations.pathmorphing,decorations.markings}

\newcommand{\Photo}{}

\begin{document}
\vspace*{4cm}
\title{Treatment of QED corrections in jet production in deep inelastic scattering at ZEUS\,\footnote{Contribution to the 2024 QCD session of the 58th Rencontres de Moriond}}
\author{Florian Lorkowski on behalf of the ZEUS Collaboration}
\address{University of Zurich}

\let\oldthefootnote\thefootnote
\renewcommand*{\thefootnote}{*}
\maketitle\abstracts{
A new measurement of inclusive jet production in deep inelastic scattering was recently published by the ZEUS Collaboration. This contribution presents a detailed discussion of the treatment of higher-order QED effects in this measurement. A comprehensive treatment of these effects is crucial for a more direct comparison between ever more precise measurements and theoretical calculations. The present analysis is the only measurement of jet production in deep inelastic scattering that can be compared to full NNLO QCD + NLO electroweak predictions.
}
\addtocounter{footnote}{-1}
\let\thefootnote\oldthefootnote

\section{Introduction}
Jet production in deep inelastic $e^\pm p$ scattering (DIS) is a valuable process to test perturbative QCD predictions and to constrain the parton distribution functions (PDFs) of the proton and the strong coupling, $\alpha_\text{s}$. Inclusive jet production is particularly well suited for this task, as it can be measured and calculated at high precision.

Recently, a new measurement of inclusive jet production in DIS was published by the ZEUS Collaboration.\cite{paper} This measurement uses the entire HERA II dataset, which accounts for more than $70\%$ of the entire luminosity collected at ZEUS. As such, this dataset provides significant new insights into this process and will be a valuable ingredient for precision fits of the PDFs and $\alpha_\text{s}$. To demonstrate its usefulness, a combined PDF and $\alpha_\text{s}$ fit was performed at next-to-next-to-leading order (NNLO) in QCD, leading to a competitive determination of the strong coupling of $\alpha_\text{s}(M_Z^2) = 0.1142 \pm 0.0019$.

In this contribution, another distinguishing aspect of the measurement will be highlighted: the treatment of higher-order QED effects. In NNLO QCD calculations for jet production in DIS, electroweak effects are usually treated at leading order, the so-called QED Born level.\footnote{Cross sections at QED Born level at defined as if no higher-order QED effects exist. In particular, this definition does not include collinear initial-state, collinear final-state and non-collinear photon radiation from the electron or the quark. Vertex corrections to the electron-boson vertex are similarly neglected. The running of the electromagnetic coupling is included, however.}\,\cite{nnlojet} In experimental analyses, higher-order QED effects are naturally present, such that measurement and calculation cannot be compared immediately. The standard approach to resolve this issue is to apply a `QED correction' to the measurement, that redefines the cross sections to Born level, thus making them comparable to the calculations. Since calculations including higher-order QED effects are not yet available, the present analysis uses a similar procedure. With ever-increasing experimental and theoretical precision, this approach is bound to become invalid at some point. Therefore, it is desirable to include higher-order QED effects into more comprehensive calculations, rather than to remove them from the measurements.

For previous measurements, this QED correction is applied in such a way that it cannot be undone, meaning that those datasets can only ever be compared to predictions at Born level. In the present analysis, it was ensured that the correction is easily reversible. Furthermore, cross sections are provided in an alternative definition, that should further facilitate more comprehensive comparisons, as soon as appropriate calculations become available.

Next-to-leading-order (NLO) electroweak corrections to the DIS process are known,\cite{hector} but not yet incorporated into most calculations. Most importantly for the present analysis, the available NNLO QCD predictions for inclusive jet production are also defined at QED Born level.\cite{nnlojet}

\section{Standard procedure for experimental treatment}
The standard procedure to make measurements and calculations comparable is to compute correction factors that redefine the measured cross sections to QED Born level. This is done by computing cross sections from two different Monte Carlo (MC) samples, one at Born level and one including higher-order effects. The ratio of those cross sections is taken as a correction factor. Vertex corrections can straightforwardly be treated with this method, as they merely change the cross section for certain phase space points. Radiative corrections, on the other hand, can change the kinematics of the event and therefore require a more careful treatment.

\begin{figure}[t]
    \tikzset{
        midarrow/.style={postaction={decorate,decoration={markings,mark=at position #1 with {\arrow{latex}}}}},
        boson/.style={decorate, decoration={snake,amplitude=0.75mm,segment length=2.5mm,mirror}}
    }
    \centering
    \begin{tikzpicture}[scale=0.7,semithick]
        \path[use as bounding box] (-3.05,-1.55) rectangle (3.75,1.9);
        \coordinate (in) at (-3,0);
        \coordinate (vertex) at (0,0);
        \coordinate (out) at (2.5,1.5);
        \coordinate (boson) at (1,-1.5);
        \coordinate (isr) at (-0.5,1);
        \coordinate (fsr) at (2.7,1.3);
        \draw[midarrow={0.5}] (in) -- (vertex) node[midway,above] {$k$};
        \draw[midarrow={0.5}] (vertex) -- (out) node[midway,above] {$k'$};
        \draw[boson] (vertex) -- (boson) node[midway,above right,xshift=0.125cm,yshift=-0.25cm] {$q=k-k'$};
    \end{tikzpicture}%
    \hspace*{0.5cm}%
    \begin{tikzpicture}[scale=0.7,semithick]
        \path[use as bounding box] (-3.05,-1.55) rectangle (3.75,1.9);
        \draw[midarrow={0.25},midarrow={0.75}] (in) -- (vertex) node[midway,below,xshift=-0.75cm] {$k$};
        \draw[midarrow={0.25},midarrow={0.75}] (vertex) -- (out);
        \draw[boson] (vertex) -- (boson) node[midway,above right,xshift=0.125cm,yshift=-0.25cm] {$\neq q = k-k'$};
        \draw[boson] ($(in)!0.5!(vertex)$) -- (isr);
        \draw[boson] ($(vertex)!0.5!(out)$) to[out=-10,in=-150] (fsr);
        \draw[decorate, decoration={brace,raise=0.5mm,amplitude=2mm}] ($1.4*(out)-0.4*(fsr)$) -- ($1.4*(fsr)-0.4*(out)$) node[midway,xshift=5mm,yshift=4mm] {$k'$};
    \end{tikzpicture}
    \caption{Left:~electron-boson vertex in the absence of QED radiation. The momentum of the electron beam $k$ is known, the final-state momentum $k'$ can be measured and the boson momentum $q$ can be computed. Right:~electron-boson vertex in the presence of QED radiation. The quantity $q = k-k'$ no longer corresponds exactly to the four-momentum of the boson.}
    \label{fig1}
\end{figure}
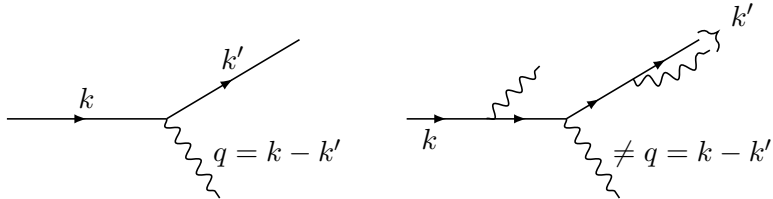

An important step towards measuring cross sections in DIS is the reconstruction of kinematic quantities, such as the momentum transfer, $Q^2$, or the Bjorken scaling variable, $x_\text{Bj}$. Multiple methods are available to do this, such as the electron method, which uses the energy and angle of the scattered DIS electron, or the double angle method, which uses the angles of the scattered electron and hadronic system. In the absence of QED radiation, these methods are equivalent up to detector resolution effects and allow accurate reconstruction of the four-momentum, $q$, of the exchanged boson. In the presence of QED radiation, especially hard collinear radiation from the initial-state electron, this is no longer the case. This is because such radiation changes the energy of the initial-state electron, which is assumed to be known in the derivation of the reconstruction methods. Since the collinear photon escapes undetected into the beam pipe, it is not straightforwardly possible to measure its energy and correct for this loss. This situation is illustrated in figure~\ref{fig1}. Since the different reconstruction methods are not equivalent, they need to be strictly distinguished and used consistently.

Soft radiation from the initial-state electron is less problematic, as it changes the momentum of the electron only marginally. Radiation from the final-state electron is usually reconstructed as part of the electron cluster and therefore does not affect the kinematic reconstruction. QED emissions from the quark are unproblematic for the reconstruction, as the quark's initial-state momentum is not used directly and final-state radiation is reconstructed as part of the jet, similarly to the electron. 

For this analysis, it was chosen to use the double angle method at detector level and the electron method at hadron level. The unfolding procedure will also correct the different choice of reconstruction method. After unfolding, an additional correction is applied that transforms the cross sections to QED Born level, where the choice of reconstruction method is no longer relevant. Schematically, the corrected cross section can be written as
\begin{equation}
    \sigma_\text{corrected} \sim
        \frac{\hat{N}_\text{Born level, MC}}{N_\text{hadron level, MC}}~\cdot~
        \frac{N_\text{hadron level, MC}}{\hat{N}_\text{detector level, MC}}~\cdot~
        \hat{N}_\text{detector level, data}, \label{eq1}
\end{equation}
where the last term represents the event count in the data, the middle term the unfolding\,\footnote{The actual analysis uses a matrix-unfolding approach, rather than the simpler bin-by-bin correction method shown here. For the current discussion, this difference is not relevant.} and the first term the QED correction. The factors that make up the first term are provided  explicitly, making it possible to undo the QED correction and obtain cross sections including higher-order effects. These are necessary to compare against more comprehensive theoretical calculations.

Choosing different reconstruction methods at detector and hadron level only has minimal impact due to one of the event selection cuts:\,\cite{paper,thesis} $38\,\text{GeV} < E-p_\text{z} < 65\,\text{GeV}$, where $E$ and $p_\text{z}$ refer to the total energy and longitudinal momentum of all reconstructed final-state particles in the event. For neutral-current DIS events, one expects $E-p_\text{z}=2(E_e - E_\text{ISR})$, where $E_e$ is the energy of the electron beam and $E_\text{ISR}$ is the energy of any potential initial-state radiation photons. Primarily, this cut helps to reject background events in which the electron is scattered at a very small angle and thus escapes undetected into the beam pipe. In addition, the cut rejects events with large amounts of initial-state radiation, as this radiation lowers $E-p_\text{z}$. For $E_e=27.5\,\text{GeV}$, the lower bound of the cut effectively translates into the cut $E_\text{ISR} < 8.5\,\text{GeV}$. Thus, the detector-level event sample does not contain hard collinear initial-state radiation, which reduces the impact of the choice of reconstruction method.

However, this cut introduces another issue. It is applied at detector level, but not at hadron level. Events with hard collinear initial-state radiation will thus be absent in the data sample, but present in the unfolded cross sections. This means that during unfolding, these events are extrapolated from the MC samples, which necessarily introduces a large dependence on the MC model. When applying the QED correction, these events are removed again, thus cancelling the model dependence. In equation~\eqref{eq1}, events counts that have hard collinear initial-state radiation removed are denoted as $\hat{N}$. The cancellation of the model dependence is evident from the structure $(\hat{N}/N \cdot N/\hat{N})_\text{MC}$ of the first and second terms. Therefore, just undoing the QED correction, as discussed above, will not allow for an optimal comparison between theory and measurement, as there will be a strong model dependence in this region of phase space.

\section{Improved treatment}
To allow for a more unbiased comparison between the measurement and comprehensive theoretical calculations, alternative correction factors are provided. These correct the unfolded cross sections not to QED Born level, but rather to a definition including higher-order QED effects and including the lower cut on $E-p_\text{z}$. These alternative cross sections can be written as
\begin{equation}
    \sigma'_\text{corrected} \sim
        \frac{\hat{N}_\text{hadron level, MC}}{N_\text{hadron level, MC}}~\cdot~
        \frac{N_\text{hadron level, MC}}{\hat{N}_\text{detector level, MC}}~\cdot~
        \hat{N}_\text{detector level, data}. \label{eq2}
\end{equation}
This definition produces cross sections including higher-order QED effects, while preserving the cancellation of model dependence. The corrected cross sections correspond closely to those events that were actually measured.

To summarise, this alternative cross section definition differs from the one given in the publication in the following regards. Firstly, it includes higher-order QED effects (vertex corrections, collinear initial-state, collinear final-state and non-collinear photon radiation from the electron or quark). Secondly, it includes the cut on hard collinear initial-state radiation $E-p_\text{z} > 38\,\text{GeV}$ or equivalently $E_\text{ISR} < 8.5\,\text{GeV}$. Thirdly, the kinematic quantities are defined according to the electron method, i.e.\ according to the right panel of figure~\ref{fig1}. This means that the quantity $q$ does not correspond exactly to the four-momentum of the exchanged boson. The difference is kept minimal by the $E-p_\text{z}$ cut.

\begin{figure}[t]
    \vspace*{-0.25cm}
    \includegraphics{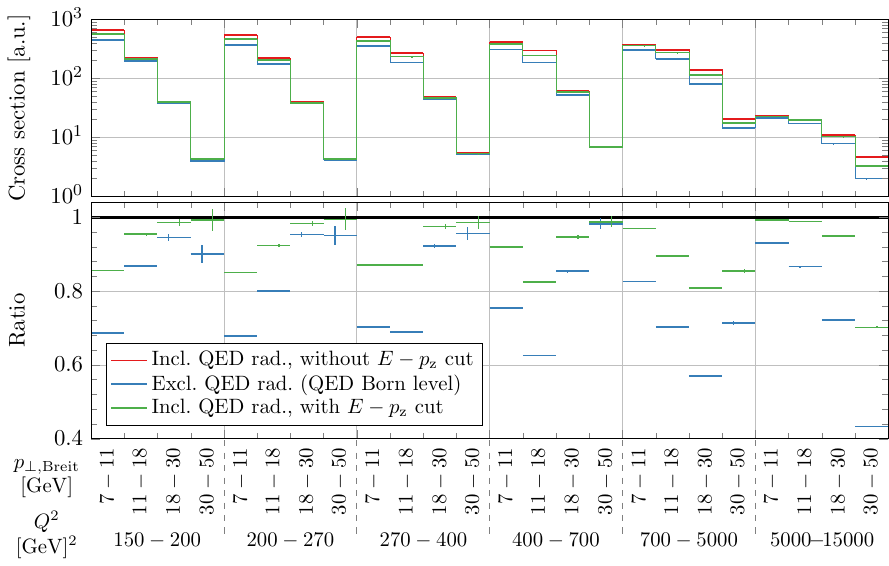}
    \vspace*{-0.25cm}
    \caption{MC predictions for the three considered cross section definitions: no QED correction (red), standard QED correction (blue) and alternative QED correction (green). The bottom panel gives the ratios of the latter two definitions to the first one. These ratios define the QED correction factors.}
    \label{fig2}
\end{figure}

The MC cross section predictions and correction factors are given in figure~\ref{fig2}.\cite{thesis} The effect of the standard QED correction is significant, often reaching about $30\%$. This is because QED radiation can affect the reconstruction of $Q^2$ and thus move events between bins of the steeply falling $Q^2$ spectrum. For the alternative QED correction, the factors are significantly smaller, usually below $15\%$. Cross sections and the correction factors are given for the standard definition in the corresponding publication\,\cite{paper} and for the alternative definition in table~C.7 of the accompanying thesis.\cite{thesis} Machine-readable versions of both tables are available on HEPData.\cite{hepdata}

\section{Summary}
A detailed explanation was given of the treatment of QED radiation in a recent inclusive jet measurement in DIS at the ZEUS experiment. Since all available calculations are performed at QED Born level, cross sections had to be determined at the same level to be comparable. In addition to that, alternative correction factors are provided that allow a straightforwardly comparison to more comprehensive calculations. The present analysis is the only measurement of jet production in DIS that can be compared to NNLO QCD + NLO electroweak calculations, when they become available.

\section*{References}
\def\Journal#1#2#3#4{{#1}\ \textbf{#2}, #3 (#4)}

\end{document}